\documentclass[twocolumn,showpacs,aps,prl,floatfix,%
superscriptaddress]{revtex4}  

\usepackage{bm} 
\usepackage{graphicx} 
 
\newcommand{\ee}{\end{equation}} 
\newcommand{\eq}{{\,=\,}} 
 
\begin{document} 
 
 
\title{Effects of jet quenching on the hydrodynamical evolution of  
quark-gluon plasma} 
\date{\today}
  
\author{A. K. Chaudhuri} 
\email[Correspond to\ ]{akc@veccal.ernet.in} 
\affiliation{Variable Energy Cyclotron Centre, 1-AF, Bidhan Nagar,  
Kolkata - 700 064, India} 
\author{Ulrich Heinz} 
\affiliation{Department of Physics, The Ohio State University,  
Columbus, OH 43210, USA} 
 
\begin{abstract} 
We study the effects of jet quenching on the hydrodynamical evolution  
of the quark-gluon plasma (QGP) fluid created in a heavy-ion collision. 
In jet quenching, a hard QCD parton, before fragmenting into a jet of  
hadrons, deposits a fraction of its energy in the medium, leading to  
suppressed production of high-$p_T$ hadrons. Assuming that the deposited 
energy quickly thermalizes, we simulate the subsequent hydrodynamic 
evolution of the QGP fluid. For partons moving at supersonic speed, 
$v_p{\,>\,}c_s$, and sufficiently large energy loss, a shock wave 
forms leading to {\em conical flow} \cite{shuryak}. 
The PHENIX Collaboration recently suggested
that observed structures in the azimuthal angle distribution 
\cite{Adler:2005ee} might be caused by conical flow. We show here 
that, for phenomenologically acceptable values of parton energy loss,
conical flow effects are too weak to explain these structures. 
\end{abstract} 
 
 
\maketitle 
 
 
Recent Au+Au collision experiments at the Relativistic Heavy Ion 
Collider (RHIC) saw a dramatic suppression of hadrons with high 
transverse momenta (``high-$p_T$ suppression'') \cite{Whitepapers}, 
and the quenching of jets in the direction opposite 
to a high-$p_T$ trigger particle \cite{STARjetqu,PHENIXjetqu}, when  
compared with p+p and d+Au collisions. This is taken
as evidence for the creation of a very dense, color opaque medium 
of deconfined quarks and gluons \cite{QGP3jetqu}. Independent  
evidence for the creation of dense, thermalized quark-gluon  
matter, yielding comparable estimates for its initial energy density 
($\langle e\rangle\agt 10$\,GeV/fm$^3$ at time 
$\tau_{\rm therm}\alt0.6$\,fm/$c$ \cite{QGP3v2}), comes
from the observation of strong elliptic flow in non-central Au+Au 
collisions \cite{Whitepapers}, consistent with ideal fluid dynamical 
behaviour of the bulk of the matter produced in these collisions.   
  
These two observations raise the question what happens, in the small  
fraction of collision events where a hard scattering produces a pair  
of high-$p_T$ partons, to the energy lost by the parton travelling  
through the medium. The STAR Collaboration has shown that, while  
in central Au+Au collisions there are no {\em hard} particles left in the  
direction opposite to a high-$p_T$ trigger particle, one sees enhanced  
production (compared to p+p) of {\em soft} (low-$p_T$) particles, broadly  
distributed over the hemisphere diametrically opposite to the trigger  
particle \cite{STARjet_therm}. This shows that the energy of the fast  
parton originally emitted in the direction opposite to the trigger 
particle is not lost, but severely degraded by interactions with 
the medium. As the impact parameter of the collisions decreases, 
the average momentum of the particles emitted opposite to 
the trigger particle approaches the mean value associated with 
{\em all} soft hadrons, i.e. the $\langle p_T\rangle$ of the 
thermalized medium \cite{STARjet_therm}. This suggests that the 
energy lost by the fast parton has been largely thermalized. 
Nevertheless, this energy is deposited locally along the fast 
parton's trajectory, leading to local energy density inhomogeneities 
which, if thermalized, should in turn evolve hydrodynamically.
This would modify the usual hydrodynamic expansion of the 
collision fireball as observed in the overwhelming number of soft 
collision events where no high-$p_T$ partons are created.  
 
Since the fast parton moves at supersonic speed,  
it was suggested in Ref.~\cite{shuryak} that a Mach shock (``sonic 
boom'') should develop, resulting in {\em conical flow} and preferred 
particle emission at a specific angle away from the direction of the 
fast parton which lost its energy. This Mach angle is sensitive to 
the medium's speed of sound $c_s$ and thus offers the possibility to 
measure one of its key properties. A recent analysis by the PHENIX 
Collaboration \cite{Adler:2005ee} of azimuthal di-hadron correlations 
in $200\,A$\,GeV Au+Au collisions revealed structures in the angular 
distribution which might be suggestive of conical flow. 

The idea of Mach shock waves travelling through compressed nuclear 
matter was first advocated 30 years ago \cite{shock,SGM74}, 
but RHIC collisions for the first time exhibit \cite{QGP3v2} 
the kind of ideal fluid behaviour which might make an extraction 
of the speed of sound conceivable. An alternate scenario, in which 
the color wake field generated by 
the fast colored parton travelling through a quark-gluon plasma 
accelerates soft colored plasma particles in the direction 
perpendicular to the wake front \cite{Stocker,RM05}, leads to an 
emission pattern which is sensitive to the propagation of 
{\em plasma} rather than {\em sound} waves \cite{RM05}. In a 
strongly coupled plasma with overdamped plasma oscillations,  
which seems to be the preferred interpretation of RHIC data 
\cite{SQCD,sQGP}, the wake field scenario should reduce to the 
hydrodynamic Mach cone picture. We here study the dynamical 
consequences of the latter, going beyond the discussion of 
linearized hydrodynamic equations in a static background 
offered in \cite{shuryak}. 
 
 
We assume that just before hydrodynamics become applicable, a pair 
of high-$p_T$ partons is produced near the surface of the fireball. 
One of them moves outward and escapes, forming the trigger jet, while 
the other enters into the fireball along, say, the $-x$ direction. 
The fireball is expanding and cooling. The ingoing parton travels 
at the speed of light and loses energy in the fireball which 
thermalizes and acts as a source of energy and momentum for 
the fireball medium. We model this medium as an ideal fluid with 
vanishing net baryon density. Its dynamics is controlled by 
the energy-momentum conservation equations 
\begin{equation} 
\label{1} 
  \partial_\mu T^{\mu\nu} =J^\nu, 
\end{equation} 
where the energy-momentum tensor has the ideal fluid form 
$T^{\mu\nu}\eq(\varepsilon{+}p)u^\mu u^\nu -p g^{\mu\nu}$, 
with energy density $\varepsilon$ and pressure $p$ being 
related by the equation of state (EOS) $p{\eq}p(\varepsilon)$,
$u^\mu\eq\gamma(1,v_x,v_y,0)$ is the fluid 4-velocity, and 
the source current $J^\nu$ is given by
\begin{eqnarray} 
\label{2} 
 &&J^\nu(x)=J(x)\,\bigl(1,-1,0,0\bigr),\\
\label{3}
 &&J(x) = \frac{dE}{dx}(x)\, \left|\frac{dx_{\rm jet}}{dt}\right| 
          \delta^3(\bm{r}-\bm{r}_{\rm jet}(t)).
\end{eqnarray} 
Massless partons have light-like 4-momentum, so the current $J^\nu$
describing the 4-momentum lost and deposited in the medium by the 
fast parton is taken to be light-like, too. $\bm{r}_{\rm jet}(t)$ is 
the trajectory of the jet moving with speed $|dx_{\rm jet}/dt|\eq{c}$.
$\frac{dE}{dx}(x)$ is the energy loss rate of the parton as it moves 
through the liquid. It depends on the fluid's local rest 
frame particle density. Taking guidance from the 
phenomenological analysis of parton energy loss observed in Au+Au 
collisions at RHIC \cite{Eloss} we take
\begin{equation}
\label{4}
  \frac{dE}{dx} = \frac{s(x)}{s_0} \left.\frac{dE}{dx}\right|_0
\end{equation}
where $s(x)$ is the local entropy density without the jet. 
The measured suppression of high-$p_T$ particle production in Au+Au 
collisions at RHIC was shown to be consistent with a parton energy 
loss of $\left.\frac{dE}{dx}\right|_0\eq14$\,GeV/fm at a reference 
entropy density of $s_0\eq140$\,fm$^{-3}$ \cite{Eloss}. For comparison, 
we also perform simulations with ten times larger energy loss, 
$\left.\frac{dE}{dx}\right|_0\eq140$\,GeV/fm.

For the hydrodynamic evolution we use AZHYDRO \cite{AZHYDRO,QGP3v2}, 
the only publicly available relativistic hydrodynamic code for 
anisotropic transverse expansion. This algorithm is formulated in 
$(\tau,x,y,\eta)$ coordinates, where $\tau{=}\sqrt{t^2{-}z^2}$ is the
longitudinal proper time, 
$\eta{=}\frac{1}{2}\ln\left[\frac{t{+}z}{t{-}z}\right]$ 
is space-time
rapidity, and $\bm{r}_\perp{\,=\,}(x,y)$ defines the plane transverse to the 
beam direction $z$. AZHYDRO employs longitudinal boost invariance 
along $z$ but this is violated by the source term (\ref{3}). We 
therefore modify the latter by replacing the $\delta$-function 
in (\ref{3}) by
\begin{eqnarray}
\label{5}
  \delta^3(\bm{r}-\bm{r}_{\rm jet}(t)) &\longrightarrow&
  \frac{1}{\tau}\,\delta(x-x_{\rm jet}(\tau))\,\delta(y-y_{\rm jet}(\tau))
\nonumber\\
  &\longrightarrow&\frac{1}{\tau} \, 
  \frac{e^{-(\bm{r}_\perp-\bm{r}_{\perp,{\rm jet}}(\tau))^2/(2\sigma^2)}}
       {2\pi\sigma^2}
\end{eqnarray}
with $\sigma{\,=\,}0.35$\,fm. Intuitively, this replaces the ``needle'' 
(jet) pushing through the medium at one point by a ``knife'' cutting the
medium along its entire length along the beam direction. The resulting
``wedge flow'' is expected to leave a stronger signal in the azimuthal 
particle distribution $dN/d\phi$ than ``conical flow'' induced 
by a single parton, since in the latter case one performs an implicit
$\phi$-average when summing over all directions of the cone normal 
vector. While a complete study of this would require a full 
(3+1)-dimensional hydrodynamic calculation, the present boost-invariant
simulation should give a robust upper limit for the expected angular
signatures. We show that the angular structures predicted from
wedge flow are too weak to explain the experimentally observed
$\phi$-modulation \cite{Adler:2005ee}.
 
The modified hydrodynamic equations in $(\tau,x,y,\eta)$ coordinates
read \cite{AZHYDRO}
%
\begin{eqnarray} 
\label{6} 
  \partial_\tau \tilde{T}^{\tau \tau} + 
  \partial_x(\tilde{v}_x \tilde{T}^{\tau \tau}) +
  \partial_y(\tilde{v}_y \tilde{T}^{\tau \tau}) 
  &=& - p + \tilde{J},
\\ 
\label{7} 
  \partial_\tau \tilde{T}^{\tau x} +
  \partial_x(v_x \tilde{T}^{\tau x}) +
  \partial_y(v_y \tilde{T}^{\tau x}) 
  &=& - \partial_x \tilde{p} - \tilde{J}, \quad
\\ 
\label{8} 
  \partial_\tau \tilde{T}^{\tau y} +
  \partial_x(v_x \tilde{T}^{\tau y}) +
  \partial_y(v_y \tilde{T}^{\tau y}) 
  &=& -\partial_y \tilde{p},  \quad
\end{eqnarray} 
%
where $\tilde{T}^{\mu\nu}\eq\tau T^{\mu\nu}$, 
$\tilde{v}_i{\eq}T^{\tau i}/T^{\tau\tau}$,
$\tilde p\eq\tau p$, and $\tilde{J}\eq\tau J$.
 
To simulate central Au+Au collisions at RHIC, we use the standard 
initialization described in \cite{QGP3v2} and provided in the 
downloaded AZHYDRO input file \cite{AZHYDRO}, corresponding to a 
peak initial energy density of $\varepsilon_0\eq30$\,GeV/fm$^3$ at 
$\tau_0\eq0.6$\,fm/$c$. We use the equation of state EOS-Q described 
in \cite{QGP3v2,AZHYDRO} incorporating a first order phase transition 
and hadronic chemical freeze-out at a critical temperature 
$T_c{\,=\,}164$\,MeV. The hadronic sector of EOS-Q is soft with a 
squared speed of sound $c_s^2 \approx 0.15$. 

%
\begin{figure}[t]
\includegraphics[bb=14 50 581 820,width=0.99\linewidth,clip]{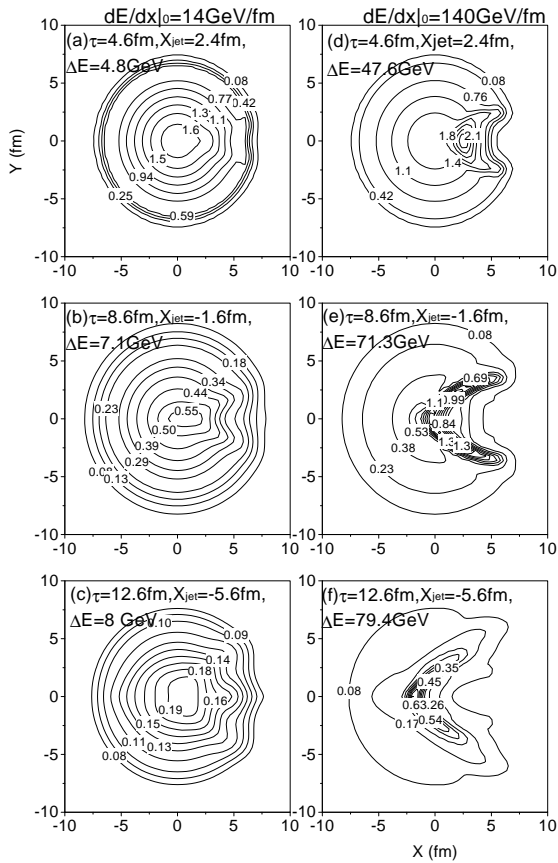}
\caption{Contours of constant local energy density in the $x$-$y$ plane 
  at three different times, $\tau{\,=\,}4.6,$ 8.6, and 12.6 fm/$c$.
  In each case the position of the fast parton, along with the integrated
  energy loss $\Delta E=\int J(x) dxdy d\tau$ up to this point, is indicated 
  at the top of the figure. Diagrams (a)-(c) in the left column were 
  calculated with a reference energy loss $dE/dx|_0{\,=\,}14$\,GeV/fm, 
  those in the right column (panels (d)-(f)) with a 10 times larger value. 
}
\label{F1}
\end{figure}
%

In our study the quenching jet starts from $x_{\rm jet}\eq6.4$\,fm
at $\tau_0\eq0.6$\,fm, moving left towards the center with constant 
speed $v_{\rm jet}\eq{c}$. For an upper limit on conical flow effects, 
the fast parton is assumed to have sufficient initial energy to emerge 
on the other side of the fireball. To simulate cases where the fast 
parton has insufficient energy to fully traverse the medium we have 
also done simulations where the parton loses all its energy within an 
(arbitrarily chosen) distance of 6.4 fm. We further compared with a 
run where the parton moved at (constant) subsonic speed 
($v_\mathrm{jet}{\,=\,}0.2\,c{\,<\,}c_s$).

\begin{figure}[t]
\includegraphics[bb=14 50 581 820,width=0.99\linewidth,clip]{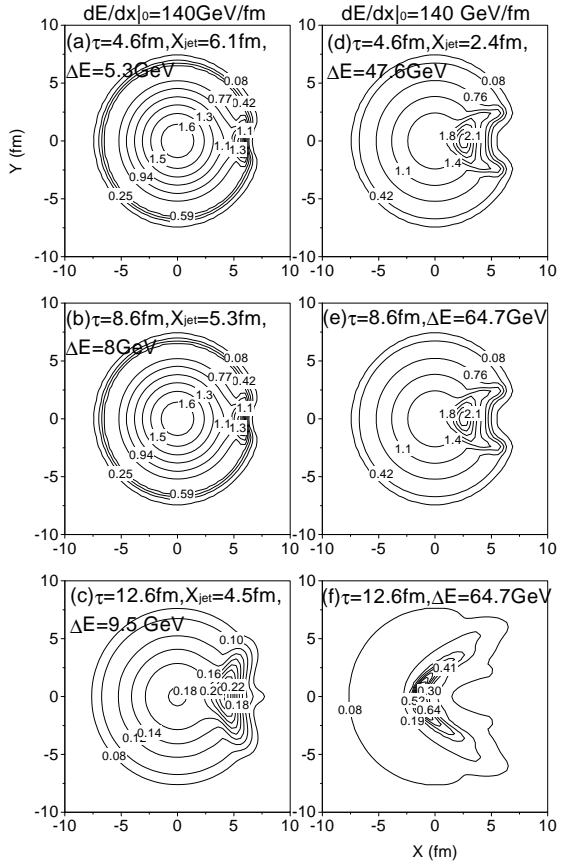}
\caption{
As in Fig.\ref{F1} but for a parton moving at subsonic speed 
$v_\mathrm{jet}{\,=\,}0.2\,c$ (left column) and for a fast parton 
($v_\mathrm{jet}{\,=\,}c$) which loses all its energy within 
the first 6.4\,fm (right column).
}
\label{F2}
\end{figure}

The resulting evolution of the energy density of the QGP fluid is 
shown in Figs.~\ref{F1} and \ref{F2}. The left column of Fig.~\ref{F1}
shows results for a phenomenologically acceptable value 
$\left.\frac{dE}{dx}\right|_0\eq14$\,GeV/fm \cite{Eloss} for the 
reference parton energy loss whereas in all other columns we use a 
ten times larger energy loss. The width of the Gaussian source
(see Eq.~(\ref{5})) is $\sigma$=0.7 fm. In the left column of
Fig.~\ref{F1} the effects of the energy deposition from the fast 
parton are hardly visible. Only for a much larger energy loss (right 
column) we recognize a clear conical flow pattern. The accumulating 
wave fronts from the expanding energy density waves build up a 
``sonic boom'' shock front which creates a Mach cone. The right columns
in Figs.~\ref{F1} and \ref{F2} show that the cone normal vector forms 
an angle $\theta_M$ with the direction of the quenching jet that is 
qualitatively consistent with expectations from the theoretical relation  
$\cos\theta_M{\eq}c_s/v_\mathrm{jet}$. However, this angle is not 
sharply defined since the cone surface curves due to 
inhomogeneity and radial expansion of the underlying medium. 
This differs from the static homogeneous case \cite{shuryak}.

%
%

When the parton travels at a subsonic speed 
$v_\mathrm{jet}{\,=\,}0.2\,c{\,<\,}c_s$ (left column in Fig.~\ref{F2}),
it doesn't get very far before the fireball freezes out due to
longitudinal expansion. In this case one only observes an accumulation 
of energy around the parton but no evidence of Mach cone formation.
When the parton travels with $v_\mathrm{jet}{\,=\,}c$ but looses all
its energy after 6.4\,fm before fully traversing the fireball (right 
column in Fig.~\ref{F2}), the fireball evolution beyond $\tau$=7\,fm 
is not affected by the fast parton directly but only indirectly through 
the propagation of earlier deposited energy. Still, Figs.~\ref{F1}f
and \ref{F2}f show that the late time evolution of the fireball is 
quite similar in both cases, demonstrating that energy deposition by
the fast parton during the late fireball stages is small, due to
dilution of the matter, and can almost be neglected.

Tests with different values for the width $\sigma$ of the Gaussian 
profile in Eq.~(\ref{5}) for the deposited energy show
that the cone angle gets better defined for smaller source size 
$\sigma$. Note that the quenching jet destroys the azimuthal symmetry 
of the initial energy density distribution but leaves the azimuthally 
symmetric energy contours to the left of the jet unaffected.

We close with a discussion of observable consequences of conical
flow. One expects \cite{shuryak} azimuthally anisotropic particle 
emission, peaking at angles $\phi\eq\pi{\,\pm\,}\theta_M$ relative to 
the trigger jet where $\theta_M$ is the Mach angle. Using the standard 
Cooper-Frye prescription, we have computed the angular distribution of 
directly emitted pions at a freeze-out temperature 
$T_\mathrm{fo}{\,=\,}100$\,MeV \cite{QGP3v2}. 
%
\begin{figure}[t] 
\includegraphics[bb=14 13 550 800,width=0.99\linewidth,clip=]%
                {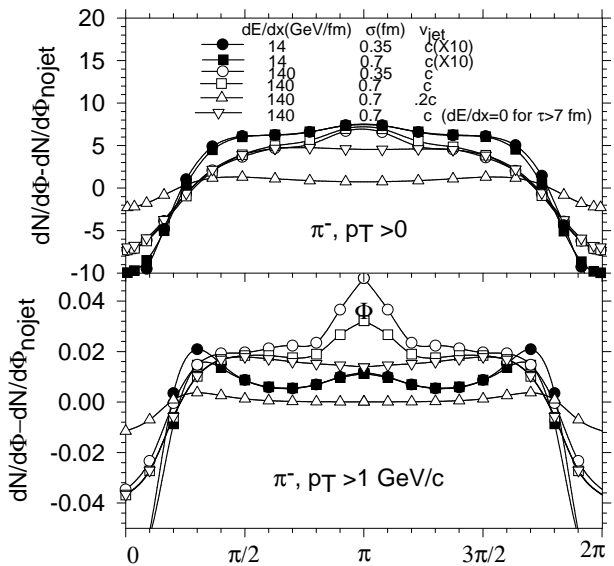}
\vspace{-4.5cm}
\caption{Azimuthal distribution $dN/dy d\phi$ of negative pions
per unit rapidity. In the upper panel we integrate over all $p_T$
while the lower panel shows only pions with $p_T{\,>\,}1$\,GeV/$c$.
Different symbols refer to different parameters as indicated. For 
better visibility the $\phi$-independent rate in the absence 
of the quenching jet has been subtracted. Filled symbols show the 
realistic case $dE/dx|_0{\,=\,}14$\,GeV/fm, enhanced by a factor 10
for visibility.
}
\label{F3}
\end{figure}
%
Figure~\ref{F3} shows the azimuthal distribution of $\pi^-$ from a 
variety of different simulations. [The $\phi$-independent constant 
$(dN_{\pi^-}/dy d\phi)_\mathrm{no\ jet}=27$ from central collisions 
without jets has been subtracted.] For $dE/dx|_0=14$\,GeV/fm the 
azimuthal modulation is very small. In none of the cases studied we 
find peaks at the predicted Mach angle with an associated dip in the 
direction of the quenched jet at $\phi\eq\pi$. One rather sees a 
{\em peak} at $\phi\eq\pi$, broadened by shoulders on both sides
which turn into small peaks at (relative to the quenching jet) 
backward angles if thermal smearing is reduced by considering only 
high-$p_T$ pions. The peak at $\phi\eq\pi$
is absent when the parton loses all its energy halfway through the 
fireball or is too slow to get to the other side before freeze-out, 
suggesting that it reflects the directed momentum imparted on the 
medium by the fast parton. It is slightly more accentuated for smaller 
$\sigma$ and higher $p_T$. We also found that the width of the shoulders 
is almost independent of the speed of sound of the medium and can not be 
used to diagnose the stiffness of its equation of state. The shoulders
exist even for subsonic parton propagation ($v_\mathrm{jet}\eq0.2\,c$, 
upright triangles), showing that other mechanisms (such as backsplash 
from the hard parton hitting the fluid and a general bias of the energy 
deposition towards the right side of the fireball due to the higher 
density of the medium at early times) have a strong influence on the 
angular distribution of the emitted particles which interferes 
with the position of the Mach peaks. The shoulder resulting from 
this combination of effects is much broader than the angular 
structures seen in the data \cite{Adler:2005ee}. The absence of 
a clear dip at $\phi\eq\pi$ in our simulations is all the more troubling
since it should have been stronger for the ``wedge flow'' studied 
here than for real ``conical'' flow. 

Our calculation does not average over the initial production points 
of the trigger particle, i.e. it ignores that in most cases its 
quenching partner does not travel right through the middle of the 
fireball cylinder, but traverses it semi-tangentially. This should 
further decrease the prominence of the shoulders in $dN/d\phi$. We 
conclude that conical flow may be able to explain the broadening
of the away-side peak in the hadron angular correlation function 
around $\phi\eq\pi$ pointed out by the STAR Collaboration 
\cite{STARjet_therm}, but is unlikely to be responsible for the 
relatively sharp structures near $\phi\eq\pi{\pm}1$ 
seen by PHENIX  \cite{Adler:2005ee}. This conclusion extends to
other conical flow phenomena, such as those generated by color wake
fields \cite{Stocker,RM05}.
 
This work was supported by the U.S. Department of Energy under contract 
DE-FG02-01ER41190.
 


\begin{thebibliography}{99} 
 
\bibitem{shuryak} 
J.~Casalderrey-Solana, E.~V.~Shuryak and D.~Teaney, 
J. Phys. Conf. Ser. {\bf 27}, 22 (2005) [hep-ph/0411315]. 
 
\bibitem{Adler:2005ee}
S.~S.~Adler {\it et al.}  [PHENIX Collaboration],
nucl-ex/0507004.

\bibitem{Whitepapers}
BRAHMS Collaboration, I. Arsene {\it et al.},  
Nucl. Phys. A {\bf 757}, 1 (2005);
%
PHOBOS Collaboration,  B. B. Back {\it et al.},  
{\it ibid.} {\bf 757}, 28 (2005);
%
STAR Collaboration, J. Adams {\it et al.}, 
{\it ibid.} {\bf 757}, 102 (2005);
%
PHENIX Collaboration, K.~Adcox {\it et al.}, 
{\it ibid.} {\bf 757}, 184 (2005).
  
\bibitem{STARjetqu} 
STAR Collaboration, C.~Adler {\it et al.}, 
Phys.\ Rev.\ Lett.\  {\bf 90}, 082302 (2003); 
STAR Collaboration, J.~Adams {\it et al.}, 
Phys.\ Rev.\ Lett.\  {\bf 91}, 072304 (2003); 
and 
Phys.\ Rev.\ Lett.\  {\bf 93}, 252301 (2004). 
 
\bibitem{PHENIXjetqu} 
PHENIX Collaboration, J.~Rak {\it et al.}, 
J.\ Phys.\ G {\bf 30}, S1309 (2004). 
 
\bibitem{QGP3jetqu} 
M. Gyulassy, I. Vitev, X.-N. Wang, and B.-W. Zhang,  
in {\it Quark-Gluon Plasma 3}, edited by R.~C. Hwa and  
X.-N. Wang (World Scientific, Singapore, 2004), p.~123. 
 
\bibitem{QGP3v2} 
P.~F. Kolb and U. Heinz, in Ref.~\cite{QGP3jetqu}, p.~634. 
 
\bibitem{STARjet_therm} 
STAR Collaboration, J.~Adams {\it et al.}, 
Phys. Rev. Lett. {\bf 95}, 152301 (2005). 
 
\bibitem{shock} 
G.~F.~Chapline, M.~H.~Johnson, E.~Teller and M.~S.~Weiss, 
Phys.\ Rev.\ D {\bf 8} (1973) 4302. 
 
\bibitem{SGM74} 
W. Scheid, H. M\"uller, and W. Greiner,  
Phys. Rev. Lett. {\bf 32}, 741 (1974). 
 
\bibitem{Stocker} 
H. St\"ocker, Nucl. Phys. A {\bf 750}, 121 (2005). 
 
\bibitem{RM05} 
J. Ruppert and B. M\"uller, Phys. Lett. B {\bf 618}, 123 (2005).
 
\bibitem{SQCD}
U. Heinz and P. F. Kolb, Nucl. Phys. A {\bf 702}, 269 (2002).

\bibitem{sQGP} 
M. Gyulassy and L. D. McLerran, Nucl. Phys. A {\bf 750}, 30 (2205); 
E. Shuryak, Nucl. Phys. A {\bf 750}, 64 (2005). 
 
\bibitem{Eloss}
X.~N.~Wang,
Phys.\ Rev.\ C {\bf 70}, 031901 (2004), 
and private communication.

\bibitem{AZHYDRO}
P. F. Kolb, J. Sollfrank, and U. Heinz, Phys. Rev. C {\bf 62}, 054909 (2000);
P. F. Kolb and R. Rapp, Phys. Rev. C {\bf 67}, 044903 (2003). The code 
can be downloaded from URL http://nt3.phys.columbia.edu/people/molnard/OSCAR/


\bibitem{AMYphot}
P.~Arnold, G.~D.~Moore and L.~G.~Yaffe,
JHEP {\bf 0112}, 009 (2001)

\end{thebibliography}
\end{document}